\documentclass[aps,prd,eqsecnum,showpacs]{revtex4}
\usepackage{graphicx}

\begin{document}
\thispagestyle{empty}
\title{Tolman-Bondi collapse in scalar-tensor theories\\
as a probe of gravitational memory}
\author{T. Harada$^{1}$\footnote{Electronic
address:harada@gravity.phys.waseda.ac.jp}, C. Goymer$^{2}$ and B. J. Carr$^{2}$}
\affiliation{
$^{1}$Department of Physics, Waseda University, Shinjuku, Tokyo
169-8555, Japan\\
$^{2}$Astronomy Unit, Queen Mary, University of London, Mile End Road,
London E1 4NS, England}
\date{\today}

\begin{abstract}
In cosmological models with a varying gravitational constant, it is not
clear whether primordial black holes preserve the value of $G$ at
their formation epoch. We investigate this question by using the
Tolman-Bondi model to study the evolution of a
background scalar field when a black hole
forms from the collapse of dust in a flat Friedmann universe.
Providing the back reaction of the scalar field on the metric can
be neglected, we find that the value of the scalar field at the event
horizon very
quickly assumes the background cosmological value. This suggests
that there is very little gravitational memory.
\end{abstract}

\maketitle

\section{Introduction}

Scalar-tensor (ST) theories of gravity provide a natural alternative to
general
relativity (GR). They describe gravity with not only a metric $g_{ab}$
but also
a scalar field $\phi$. Derivatives of $\phi$ appear as source terms in
the field
equations and $\phi$ itself satisfies a wave equation. The strength of the
gravitational coupling is determined by the function $\omega(\phi)$,
where GR is recovered in
the limits $\omega\rightarrow\infty$ and 
$\omega^{-3}(d\omega/d\phi)\rightarrow0$.

ST theories can also be regarded as being equivalent to GR with a varying
gravitational ``constant'' $G$. The most simple example of such a ST
theory is
Brans-Dicke theory \cite{BD}, where $\omega(\phi)$ is constant and
$G\propto\phi^{-1}$. However,
weak field experiments have shown that $\omega>500$ \cite{Will} and so the
deviation from GR is small. For more general ST theories, where $\omega$
is not
constant, it is possible that $\omega$ was much smaller at earlier
times. So observations allow such theories to greatly deviate from GR
in the early universe.

There has been a renewed interest in ST theories in recent years due to the
effective low energy actions of string theory involving one or more scalar
fields. These scalar fields enter the field equations in much the same way
as the scalar field in ST theories \cite{Call}. Also, the increasing
popularity of
inflation and quintessence suggests that scalar fields might
need to be incorporated into
cosmological models.

The purpose of this paper is to study the effect of an evolving scalar
field on the formation and evolution of
a primordial black hole. In an asymptotically flat spacetime it is well
known that a
black hole radiates away any inhomogeneities in the scalar field until
it becomes a stationary
solution with constant $\phi$ \cite{Hawk}. This is a consequence of the
famous ``no hair'' theorem. However, in ST cosmological models the
scalar field
is evolving with time and this would modify how the black hole evolves
during
its lifetime.

Barrow \cite{Barr} was the first to examine this problem. He considered the
two extreme possibilities: scenario (A), where the scalar field
evolves
everywhere homogeneously in the same way as the cosmological background;
and scenario
(B), where the black hole forces the scalar field to remain constant in
some
local region around it. The second scenario Barrow called gravitational
memory because the black hole would locally preserve the value of the
scalar
field from when it formed. Barrow \& Carr \cite{BC} studied the evolution
of primordial black holes for these scenarios and found that either case
results in a
significant deviation from the usual GR analysis. They considered ST
theories,
where $G(\phi)\propto\phi^{-1}$ as in the Brans-Dicke case. Since $\phi$
increases with cosmological time, this implies that $G(\phi)$ decreases, so
black holes would take longer to radiate away their mass via Hawking
evaporation than in the GR case. In both scenarios the black holes form
when
gravity is stronger, so the rate of evaporation is less, but in
scenario (B) the strength never decreases and so the lifetime is even
longer.

The above scenarios are the two extremes and the
reality is
probably somewhere in between. Two more general scenarios have been
proposed
\cite{GC}. In scenario (C) the
scalar field evolves faster at the event horizon (EH) than at the particle
horizon (PH), so $\dot{\phi}_{EH}>\dot{\phi}_{PH}$. Eventually the
black hole must reach a stage where
$\dot{\phi}_{EH}=\dot{\phi}_{PH}$, but this does not necessarily mean the
scalar field is homogeneous since there could be some lag between the
asymptotic and local increase. In scenario (D), the scalar field is
evolving locally
but at a
slower rate than asymptotically, so
$\dot{\phi}_{EH}<\dot{\phi}_{PH}$. There is still some gravitational
memory but not in the
strict sense of scenario (B). In this scenario the gradient of the scalar
field is increasing but one would expect there to be some limit,
due to the influx of scalar gravitational waves.

Gravitational memory for black holes which are small compared to the
cosmological scale (i.e. the particle
horizon) has already been investigated by Jacobson
\cite{Jaco}. In this case, the scalar field evolution can be considered
as an asymptotic
perturbation to the Schwarzschild metric and the end-state of scenario (C) applies, with the lag being found to be small. This suggests that gravitational memory is
virtually
non-existent. However, this approximation may not apply for
primordial black holes since these can have a size comparable to the
particle
horizon at formation \cite{CH}. It is still not clear what would happen in this case.
Therefore another way of investigating gravitational memory, without
assuming that
the black hole is small, is needed.

In GR there have been several attempts to study analytic solutions which
represent
black holes within a cosmological background. The earliest used the
Einstein-Straus solution \cite{ES}, which matches a Schwarzschild interior to a
Friedmann
exterior, and this approach has also been used to study gravitational memory
\cite{Sak}. However, in most circumstances it can be shown \cite{Whin} that such a matching is only
possible if the scalar field is constant, which just gives the GR solution.
Another method used the McVittie metric \cite{McV}, but it has been shown
\cite{Nola} that this has a scalar curvature singularity at the event
horizon.

A more successful method uses the Tolman-Bondi metric \cite{TB} to represent the
collapse of dust to a black hole in an asymptotically Friedmann background.
However, this only works for dust and cannot in general be applied
to ST theories due to the derivatives of the scalar field
appearing as source terms in the field equations.
In this paper we overcome this problem by assuming that the
effect of the scalar field on the spacetime is small compared to that of
the matter. This means that we can use the usual Einstein field equations
to generate the spacetime and then use the wave equation for the scalar
field to determine its evolution. This approximation was used by Harada
et al.
\cite{HCNN} to calculate the scalar gravitational radiation emitted by
Oppenheimer-Snyder collapse in ST theory. Jacobson also uses this
approximation when calculating the effect of an evolving scalar field in
Schwarzschild spacetime. We note, however, that
self-consistent numerical calculations of spherical 
gravitational collapse
in ST theory in asymptotic flat spacetime,
in which the effect of the scalar field on the spacetime
is fully incorporated, have been considered by previous authors \cite{SNN,SST}.

If one makes this approximation to investigate gravitational collapse
in a Tolman-Bondi spacetime, the solution is specified by two arbitrary
functions: the
energy and mass functions. To represent collapse in a
Friedmann background, the energy function has to be negative
within some radius $r_0$ and zero outside it. This results in the eventual
gravitational collapse of all the matter within $r_0$, while the matter
outside $r_0$
expands forever as in a flat Friedmann model. In choosing the mass
function, or equivalently
the density perturbation, we adopt the ``compensated" model. This means
that the overdense region in the centre
is surrounded by an underdense region outside, so that the total mass at
infinity is unaffected.

We solve the field equations numerically using the characteristic method.
This method was first applied to an inhomogeneous and
dynamical background spacetime by Iguchi
et al. \cite{IHN}. The characteristic method integrates over null
hypersurfaces
with the event horizon as a boundary. This means that one never needs to
calculate anything inside the black hole, thereby avoiding any numerical
problems associated with singularities.
The output of the code shows the spatial and temporal variation of the scalar field. The
figures produced
show that the initial collapse results in a large gradient in the scalar
field.
However, as time increases, the scalar field becomes almost homogeneous.
This
suggests that, within the approximation used, gravitational
memory is not possible. It remains to be seen whether
the back reaction of the scalar field could alter this conclusion.

In Section II we describe ST theories in more detail and derive the
field equations
for the approximation in which the effect of the scalar field on the
metric is neglected. In Section III we transform the Tolman-Bondi
solution into null coordinates, giving
the equations necessary to apply the characteristic method. In Section
IV we specify
the model giving rise to black hole formation. We present the
numerical results in Section V and discuss their implications in Section VI.

\section{Basic equations}

For scalar-tensor theories of gravity with $G(\phi)\propto\phi^{-1}$ the
field
equations are

\begin{equation}
\label{field1}
G_{ab} = \frac{8\pi}{\phi} T_{ab} + \frac{\omega(\phi)}{\phi^{2}}
\left( \partial_a\phi \partial_a\phi - \frac{1}{2} g_{ab} \phi^c \phi_c
\right)
+ \frac{1}{\phi}(\nabla_a \nabla_b \phi - g_{ab} \nabla^c \nabla_c \phi),
\end{equation}

\begin{equation}
\label{field2}
\nabla^c \nabla_c \phi = \frac{8\pi T - (d\omega/d\phi)
\partial^c\phi \partial_c\phi} {3 + 2\omega(\phi)}
\end{equation}
where $T_{ab}$ is the usual energy-momentum tensor, $T$ is its trace and
we have set $c=1$. In Brans-Dicke theory Eq.(\ref{field1}) remains
unchanged but
$d\omega/d\phi=0$ in Eq.(\ref{field2}). These equations are
expressed in the Jordan frame but ST theories can also be expressed in
a conformal frame known as the Einstein frame. The Einstein frame is
related to the Jordan
frame by the transformation

\begin{equation}
\bar{g}_{ab}= (G_0 \phi)g_{ab} \quad \Rightarrow \quad \bar{T}_{ab}=
(G_0 \phi)^{-1} T_{ab},
\end{equation}
where $G_0$ is the present value of the gravitational ``constant'' as
measured
in solar system experiments. It is called the Einstein frame because it
can be
expressed as GR with a scalar field. However, the Jordan frame will be used
throughout this paper.

In GR the Tolman-Bondi solution is given by

\begin{equation}
ds^{2}=-dt^{2}+A^{2}(t,r)dr^{2}+R^{2}(t,r)[d\theta^{2} +\mathrm{sin}^2
\theta d\psi^{2}],
\end{equation}
where $r$ is the comoving radial coordinate and $R$ is given by
\begin{equation}
t-t_{s}(r)=\sqrt{\frac{R^{3}}{F}} G\left(-\frac{fR}{F}\right).
\label{evolution}
\end{equation}
Here $G(y)$ is a positive function given by
\begin{equation}
G(y)= \left\{
\begin{array}{ll}
\displaystyle{\frac{\mbox{Arcsin}\sqrt{y}}{y^{3/2}}}-
\displaystyle{\frac{\sqrt{1-y}}{y}}\qquad \mbox{or}\qquad
\displaystyle{\frac{\pi-\mbox{Arcsin}\sqrt{y}}{y^{3/2}}}+
\displaystyle{\frac{\sqrt{1-y}}{y}}&\qquad (0<y\le 1) \\
\displaystyle{\frac{2}{3}}&\qquad (y=0) \\
\displaystyle{\frac{-\mbox{Arcsinh}\sqrt{-y}}{(-y)^{3/2}}}-
\displaystyle{\frac{\sqrt{1-y}}{y}}&\qquad (y<0)
\end{array}\right.,
\end{equation}
$t_s(r)$ is a constant of integration, and $F(r)= 2G_0 m(r)$ with $m(r)$ being
the mass within radius $r$. $A$ is given by
\begin{equation}
A^{2}(t,r)=\frac{R^{\prime}(t,r)^2}{1+f(r)},
\end{equation}
and $R$ satisfies
\begin{equation}
\dot{R}^2(t,r)=\frac{F(r)}{R}+f(r).
\label{eq:energy}
\end{equation}
The density of the dust $\rho$ is given by
\begin{equation}
\rho=\frac{F^{\prime}}{8\pi R^{2}R^{\prime}}.
\end{equation}
In the above,
a dot denotes $ \partial_{t}$ and a prime denotes $ \partial_{r}$.

There are two arbitrary functions in this solution: the mass function
$m(r)$ and the energy function $f(r)$. Investigating collapse to a black
hole just requires the appropriate choice for these
functions.
In this paper the energy function is chosen such that

\begin{equation}
f(r)<0 \quad {\rm for} \quad r<r_0
\end{equation}
\begin{equation}
f(r) = 0 \quad {\rm for} \quad r > r_0
\end{equation}
for some $r_0$. This means that when a perturbation is applied to the
background dust, all the matter
interior to $r_0$ will eventually collapse to form a black hole, while the
exterior matter will expand forever as in a flat Friedman universe
\cite{DL}.
The mass function $m(r)$ is determined by putting $R=r$ at $t=t_s(r)$ in 
Eq.~(\ref{evolution}).

The key to this approximation is that the back reaction of the scalar field
is neglected. It is assumed that the effect of the scalar field on the
spacetime
is small compared to that of the matter. The initial configuration used
is the
general relativistic one with constant $\phi$ and for simplicity
Brans-Dicke theory is used. Then, to the lowest order, the evolution of
$\phi$
is determined by the wave equation:

\begin{equation}
\left[\nabla^c\nabla_c\right]_{TB}\phi = \frac{8\pi}{3+2\omega} T_{TB}
\label{eq:scalar}
\end{equation}
where $\left[\nabla^c\nabla_c\right]_{TB}$ and $T_{TB}$ are determined
for the Tolman-Bondi metric and the general relativistic solution. Using
the Tolman-Bondi metric,  the wave operator is given by

\begin{equation}
\left[\nabla^c\nabla_c\right]_{TB}\phi =-\ddot{\phi}
-\frac{(AR^{2})^{\cdot}}{AR^{2}}\dot{\phi}
+\frac{1}{A^{2}}\phi^{\prime\prime}+\frac{1}{AR^{2}}\left(
\frac{R^{2}}{A}\right)^{\prime}\phi^{\prime}.
\end{equation}

\section{Characteristic method}

The last equation needs to be rewritten in terms of a null coordinate
suitable for
the characteristic method. The retarded time coordinate $u$ is
introduced such
that $u=\mathrm{constant}$ is an outgoing null geodesic. In the original
coordinates the outgoing null geodesic is given by
\begin{equation}
\frac{dt}{dr}=A,
\end{equation}
so we can write
\begin{equation}
\frac{u^{\prime}}{\dot{u}}=-A.
\end{equation}
The coordinate system is now transformed from $(t,r)$ to $\left(u(t,r),
\bar{r}(r)\right)$ using the relations

\begin{equation}
du=\frac{1}{\alpha}(dt-Adr), \quad d\bar{r}=dr,
\end{equation}
where
\begin{equation}
\alpha\equiv\frac{1}{\dot{u}} .
\end{equation}
The partial derivatives are then related
by
\begin{equation}
\partial_{t}=\frac{1}{\alpha}\partial_{u}, \quad
\partial_{r}=-\frac{A}{\alpha}\partial_{u}+\partial_{\bar{r}}.
\end{equation}
In this coordinate system the metric becomes
\begin{equation}
ds^{2}=-\alpha^{2} du^{2}-2\alpha A(u,\bar{r}) du d\bar{r}+ R^{2}(u,\bar{r})
[d\theta^{2}+\mathrm{sin}^2\theta d\psi^{2}].
\end{equation}

To use the characteristic method it is necessary to introduce the
derivative
along the ingoing radial null geodesic. In the original coordinates the
ingoing
null geodesic is given by

\begin{equation}
\frac{dt}{dr}=-A,
\end{equation}
which in the new coordinate system becomes

\begin{equation}
\frac{d\bar{r}}{du}=-\frac{\alpha}{2A}.
\end{equation}
Therefore the derivative along the ingoing null can be obtained:

\begin{equation}
\frac{d}{du}=\partial_{u}+\frac{d\bar{r}}{du}\partial_{\bar{r}}
=\frac{\alpha}{2}\left(\partial_{t}-\frac{1}{A}\partial_{r}\right).
\end{equation}
The partial derivatives $\partial_t$ and $\partial_r$ can now be rewritten
in terms of $d/du$ and $\partial_{\bar{r}}$:

\begin{eqnarray}
\label{partialt}
\partial_{t}&=&\frac{1}{\alpha}\frac{d}{du}+\frac{1}{2}\frac{1}{A}
\partial_{\bar{r}}, \\
\label{partialr}
\partial_{r}&=&-\frac{A}{\alpha}\frac{d}{du}+\frac{1}{2}\partial_{\bar{r}}.
\end{eqnarray}
The wave operator then becomes

\begin{equation}
\left[\nabla^c \nabla_c\right]\phi = -\frac{2}{\alpha AR}\frac{d\varphi}{du}
-\frac{A^{\prime}}{A^{3} R}\varphi
+\frac{1}{AR}\left[(A\dot{R})^{\cdot}-\left(\frac{R^{\prime}}{A}\right)
^{\prime}\right]\phi\\
\end{equation}
where
\begin{equation}
\varphi \equiv \partial_{\bar{r}}(R\phi).
\end{equation}
Here $^{\cdot}$ and
$^{\prime}$ refer to the operators given in Eq.(\ref{partialt}) and
Eq.(\ref{partialr}) respectively.
It is also necessary to obtain an equation for $\alpha$. This is
achieved by
using $(\dot{u})^{\prime}=(u^{\prime})^{\cdot}$, which gives

\begin{equation}
\partial_{\bar{r}}\alpha=\dot{A}\alpha.
\end{equation}

Applying the full Tolman-Bondi solution, the basic equations that must
be solved in Brans-Dicke theory are to first order

\begin{eqnarray}
\frac{d\varphi}{du}&=&\frac{\alpha}{2}\frac{\sqrt{1+f}}{R^{\prime}}
\left(\frac{f^{\prime}}{2(1+f)}-\frac{R^{\prime\prime}}{R^{\prime}}\right)
\varphi
+\frac{\alpha}{2}\frac{F}{R\sqrt{1+f}}\left(\frac{F^{\prime}}{2F}
-\frac{R^{\prime}}{R}\right)\phi \nonumber \\
& &+\frac{\alpha}{2}\frac{1}{3+2\omega}
\frac{F^{\prime}}{R\sqrt{1+f}}, \\
\partial_{\bar{r}}\alpha&=&\pm\frac{\alpha}{2}
\frac{1}{\sqrt{(1+f)\left(\frac{F}{R}+f\right)}}
\left(\frac{F^{\prime}}{R}-\frac{FR^{\prime}}{R^{2}}+f^{\prime}
\right),
\end{eqnarray}
where the upper and lower signs correspond to an expanding and
collapsing phase respectively.
For numerical purposes it is also convenient to take the parametrized form
of the Tolman-Bondi solution.
For $f=0$, $R$ is given by
\begin{equation}
R=\left(\frac{9F}{4}\right)^{1/3}(t-t_{s}(r))^{2/3}.
\end{equation}
For $f>0$, it is given by
\begin{eqnarray}
R&=&\frac{F}{2f}(\cosh\eta-1), \\
t-t_{s}(r)&=&\frac{F}{2f^{3/2}}(\sinh\eta-\eta).
\end{eqnarray}
For $f<0$, it is given by
\begin{eqnarray}
R&=& \frac{F}{2(-f)}(1-\cos\eta), \\
t-t_{s}(r)&=&\frac{F}{2(-f)^{3/2}}(\eta-\sin\eta).
\end{eqnarray}
Here the signs are chosen
so that they correspond to the big bang universe.

\section{Models}

We choose the background primordial black hole model
so that the following conditions are satisfied.
(1) The big bang occurs at the same time everywhere, i.e. 
$t_s(r)=0$ (constant).
(2) The model is asymptotically flat Friedmann and compensated (i.e. the overdense region in the centre
is surrounded by an underdense region outside in such a way that the total mass at
infinity is unaffected).
(3) The model is free of shell-focusing or
shell-crossing naked singularities,
at least within the calculated region.
(4) The central region is bound, while the
asymptotic region is marginally bound.
(5) At the initial time $t=t_{0}$, the condition $R'>0$ is satisfied
everywhere.

In order to satisfy the above conditions, we set 
$t=t_{0}$ and
choose the energy function $f(r)$ to have the form
\begin{equation}
f(r)=\left\{
\begin{array}{cl}
-\left(\displaystyle{\frac{r}{r_{\rm c}}}\right)^{2} & 
\mbox{for}\quad r<r_{\rm w} \\
-\left(\displaystyle{\frac{r}{r_{\rm c}}}\right)^{2}\exp
\left(-\left(\displaystyle{\frac{r-r_{\rm w}}
{r_{\rm w}}}\right)^{4}\right)
& \mbox{for}\quad r\ge r_{\rm w}
\end{array}\right.
\label{eq:f}
\end{equation}
where $r_{\rm c}$ gives the curvature radius in the central closed
Friedmann region and $r_{\rm w}$ gives the scale of the
overdense region. 
Equation~(\ref{eq:f}) means that the central region $r<r_{\rm w}$
is described by the exact closed Friedmann solution,
which ensures that there is no shell-focusing naked singularity. 
More general situations in which shell-focusing is avoided
have been discussed elsewhere \cite{Josh}. 
The code continually monitors for shell-crossing to 
check that this never happens within the calculated region.
We then determine $F(r)$ so that $r$ coincides with $R$ at the
$t=t_{0}$ spacelike hypersurface.

The choice of the function $f(r)$ requires some justification since it
is nowhere exactly zero in the calculated region, so strictly speaking
the entire region would eventually collapse to a black hole if one
waited long enough. In practice, this does not matter because our
calculated region is so large that the value of $f$ is effectively
identical to $0$ for $r\agt 5r_{\rm w}$ (i.e. at the outer
boundary). Since the region is finite, we can therefore always make the
matching to the Einstein-de Sitter universe {\it outside} the calculated
region. It would be easy in principle to perform the calculations for a
situation in which the matching occurs {\it within} the calculated
region  (i.e. with $f=0$ exactly at the edge of the region). It is
clear that this would make no qualitative difference to our
conclusions but, to confirm this,
we also adopt the choice
\begin{equation}
f(r)=\left\{
\begin{array}{cl}
-\left(\displaystyle{\frac{r}{r_{\rm c}}}\right)^{2}
\left[1-\left(\displaystyle{\frac{r}{r_{w}}}\right)^{4}\right]^{4} & 
\mbox{for}\quad r<r_{\rm w} \\
0 & \mbox{for}\quad r\ge r_{\rm w}
\end{array}\right.,
\label{eq:fpol}
\end{equation}
in which the compensation is satisfied explicitly within the 
calculated region.

Before integrating,
we have to fix the initial data for $\phi$.
One can take this to be
the homogeneous cosmological
solution given by
\begin{equation}
\phi_{\rm c}=\phi_{0}\left(1+\frac{1}{3+2\omega}\frac{4}{3}
\ln \frac{t}{t_{0}}\right).
\label{eq:cosmological}
\end{equation}
We then set the initial null hypersurface as the null cone
whose vertex is at $(t,r)=(t_{0},0)$
and regard the cosmological solution as the initial
data on this
hypersurface.
Although the value of the scalar field at the
cosmological particle horizon must be given
by this solution, the value in the perturbed region
and the surrounding region
may be different from this.
To examine the sensitivity of the results to this alteration,
we consider another form of the initial data
which is different from the cosmological solution
in the region $t \alt (1-10)\times t_{0}$. We now choose
\begin{equation}
\phi_{\pm}=\phi_{\rm c}\left[1\pm \exp
\left[-\left(\frac{t}{5 t_{0}}\right)^{2}\right]\right],
\end{equation}
so that we have an ingoing wave in the perturbed region,
and examine the evolution of the scalar field thereafter.
The numerical code has been checked by the following non-trivial
test calculation. In the flat Friedmann universe,
the code must reproduce the cosmological evolution
Eq.~(\ref{eq:cosmological})
from the initial data.
There is agreement to within 0.05\% accuracy.

\section{Results}

We denote the Hubble parameter in the Friedmann background
(far from the perturbed region) as $H_{0}$ at $t=t_{0}$. Recall that
$r_{\rm c}$ gives the amplitude of the density perturbation, while
$r_{\rm w}$ gives the size of the perturbed region.
For super-horizon scale perturbations, we cannot set
the density perturbation to be very large else
the overdense region closes up on itself and
becomes disconnected from the rest of the universe \cite{CH}.
Actually, the requirement $R'>0$ imposes an even stronger condition since
it implies $r_{\rm w}\alt r_{\rm c}$.
We have set the Brans-Dicke parameter to be $\omega=5$.
If $r_{\rm c}$ is much increased, then the amplitude of the overdensity
is much decreased and the resulting black hole becomes
very small compared
with the horizon scale at the formation time.
If $r_{\rm c}$ is much decreased, the overdense region becomes a separate
closed universe.
Models and parameters are summarized in Table~\ref{tb:models}.
The difference between models A, B, C and F is in the choice of the
background perturbation.
The difference between models A, D and E is in the
choice of the initial condition for the scalar field.
A change of $\omega$ only scales the variation of the scalar field
from $\phi_{0}$, as indicated by Eq.~(\ref{eq:cosmological}).

In Fig.~\ref{fg:fd},
the energy function $f(r)$ and the initial density perturbation
\begin{equation}
\delta(t_{0},r)\equiv
\frac{\rho(t_{0},r)-\rho(t_{0},\infty)}{\rho(t_{0},\infty)}
\end{equation}
are plotted.
The trajectories of outgoing null geodesics
are plotted in Fig.~\ref{fg:null}.
It is seen that a nearly horizon-scale black hole
is formed for model A, while the black hole
is smaller than the horizon scale for models B, C and F.
This is because the size of
the black hole is always roughly $20H_0^{-1}$, whereas the time at which it forms  is about $20H_0^{-1}$ in model A and closer to $100H_0^{-1}$ in the other cases. The initial data for the scalar field are set on the initial
null Cauchy surface (see Fig.~\ref{fg:penrose}).
We prepare three sets of initial data, $\phi_{\rm c}$, $\phi_{+}$
and $\phi_{-}$, and these are plotted in Fig.~\ref{fg:initialdata}.
We have investigated all the models listed in Table~\ref{tb:models} and
the results are seen in
Figs.~\ref{fg:tp}-\ref{fg:horizon}. In Fig.~\ref{fg:tp},
the profile of the scalar field
is plotted for constant $t$.
In Fig.~\ref{fg:rp},
it is plotted for
constant $R$.
The reason why some of the curves come to an abrupt end 
{\it below} some value of $R$ is that the event horizon has formed
near the center and we did not calculate the evolution
of the scalar field inside the event horizon. The reason why some of 
them come to an abrupt end 
{\it above} some value of $R$ is due to the finiteness of the region in which the numerical calculation is done.

We note that collapse ensures that the scalar field is initially concentrated in the central regions and this means that it rises above the asymptotic cosmological value everywhere. However, this central concentration tends to fall due to the underdensity surrounding the black hole. This effect, coupled with the increase of the cosmological value, means that the scalar field necessarily falls below the cosmological value at sufficiently large values of $R$, at least for the models under consideration. Eventually it may do so at every plotted value of $R$. Strictly speaking, the issue of gravitational memory is concerned with the process whereby the scalar field is {\it raised} to the cosmological value once it
has fallen below it rather than with the process whereby it initially {\it falls} to the cosmological value. 

Generally, the configuration of the scalar field tends
to spatially homogeneity with increasing time and the value of the scalar
field around the black hole follows the cosmological evolution
of the scalar field at each moment.
To see this more clearly, we identify the value of the scalar field
on the latest null ray within the calculation
as the value on the black hole
event horizon, $\phi_{\rm BH}$,
because the null ray is very close
to the event horizon
for each model (at least for $H_{0}t \alt 300$, see Fig.~\ref{fg:null}).
Both $\phi_{\rm c}$ and $\phi_{\rm BH}$ are plotted
in Fig.~\ref{fg:horizon}.
The end of the curve $\phi_{\rm BH}$ corresponds to the formation time
of the event horizon. 
We can see that $\phi_{\rm BH}$ follows the evolution of $\phi_{\rm c}$.
There is a small deviation of $\phi_{\rm BH}$ from $\phi_{\rm c}$ but this
can be explained by the central overdensity and
the surrounding underdensity.
For $H_{0}t \alt 100$, the event horizon runs through
the overdense region
and hence the scalar field is amplified compared with $\phi_{\rm c}$.
Thereafter the event horizon runs through the underdense region
and so the scalar field is reduced compared with $\phi_{\rm c}$.
The overall evolution of $\phi_{\rm BH}$ is well described
by $\phi_{\rm c}$.

Although
there are minor differences among the models, we can conclude that
the configuration of the scalar field is nearly spatially homogeneous
and well described by the cosmological solution $\phi_{\rm c}$,
at least for around ten initial Hubble times after the formation of
the event horizon.

\section{Summary}
We have calculated the evolution of
the Brans-Dicke scalar field in the
presence of a primordial black hole
formed in a flat Friedmann background.
We have found that
the value of the scalar field
at the event horizon almost always maintains
the cosmological value.
This suggests that
primordial black holes ``forget''
the value of the gravitational constant
at their formation epoch. In this sense, we confirm the result of Jacobsen \cite{Jaco}, although he never carried out an explicit evolutionary calculation.
While it is possible that some choices of the function $f(r)$ might lead to a different conclusion, we would claim that our conclusion must at least hold in compensated models. For we have tried many sets of parameters, corresponding to models which describe a wide variety of physical situations, and this always seems to be the case. However, it is not clear from our
analysis what would happen for non-compensated models and we have also avoided models with shell-crossing singularities, so we do not claim that our result is completely general.

It should also be stressed that this result has only been
demonstrated for a dust universe in which the scalar
field does not appreciably affect the background curvature and
it remains to be seen whether the same conclusion applies when this
assumption is dropped. As can be seen from Figs. \ref{fg:tp}
and \ref{fg:rp}, both the radial
gradient and the time derivative of $\phi$ are large at early times
and both of these act as source terms in the field equations. However,
as long
as $\omega$ is large, this is unlikely to stop the scalar field becoming
homogeneous eventually, since it is clear that both the radial gradient and
time derivative become small well after the initial collapse. Neglecting
the back reaction should therefore be
a reasonable assumption in this situation, so gravitational memory
seems unlikely. However, this conclusion might
not apply for $\omega \approx 1$.

\section*{Acknowledgments}
We would like to thank T.~Nakamura for helpful discussion.
TH is supported by the Grant-in-Aid (No. 05540)
from the Japanese Ministry of
Education, Culture, Sports, Science and Technology. CG
is supported by the UK Particle Physics \& Astronomy Research Council.

\begin{table}[htbp]
\begin{center}
\caption{Parameters for models}
\label{tb:models}
\begin{tabular}{c|cccc}
Models & $f$ & $H_{0}r_{\rm c}$ & $H_{0}r_{\rm w}$ & initial data \\ \hline
A & (\ref{eq:f})& 2 & 1.25 & $\phi_{\rm c}$ \\
B & (\ref{eq:f})& 2 & 1 & $\phi_{\rm c}$ \\
C & (\ref{eq:f})& 3 & 1.25 & $\phi_{\rm c}$ \\
D & (\ref{eq:f})& 2 & 1.25 & $\phi_{+}$ \\
E & (\ref{eq:f})& 2 & 1.25 & $\phi_{-}$ \\
F & (\ref{eq:fpol}) & 2 & 3 & $\phi_{c}$ \\
\end{tabular}
\end{center}
\end{table}

\begin{figure}[htbp]
\caption{(a) Energy function $f$ and (b)
initial density perturbation $\delta$ are plotted for models A-F.}
\label{fg:fd}
\caption{Trajectories of outgoing null rays are plotted (a) for models
A,D,E, (b) for model B, (c) for model C and (d) for model F.}
\label{fg:null}
\caption{Penrose diagram of a primordial black hole in a flat Friedmann universe. Also depicted is the null Cauchy surface on which
the initial conditions are set for the numerical calculations.}
\label{fg:penrose}
\caption{Initial null Cauchy data sets for the scalar field: $\phi_{\rm c}$ for
models A,B,C and F; $\phi_{+}$ for model D; and $\phi_{-}$ for model E.
For clarity, the abscissa is plotted logarithmically.}
\label{fg:initialdata}
\caption{Configuration of the scalar field
at each moment $t=\mbox{const}$ is plotted
for models (a) A, (b) B, (c) C, (d) D, (e) E and (f) F.
For comparison, the cosmological value $\phi_{\rm c}$
at each moment is also plotted as a horizontal line.}
\label{fg:tp}
\caption{Time variation of the scalar field
along the world-lines of constant $R$
is plotted for models
(a) A, (b) B, (c) C, (d) D, (e) E and (f) F.
For comparison, the cosmological evolution $\phi_{\rm c}$ is
also plotted.}
\label{fg:rp}
\caption{Time evolution of the scalar field
on the event horizon is plotted for models
(a) A, (b) B, (c) C, (d) D, (e) E and (f) F.
For comparison, the cosmological evolution $\phi_{\rm c}$ is
also plotted.}
\label{fg:horizon}
\end{figure}
\begin{figure}[htbp]
(a)\includegraphics[scale=0.5]{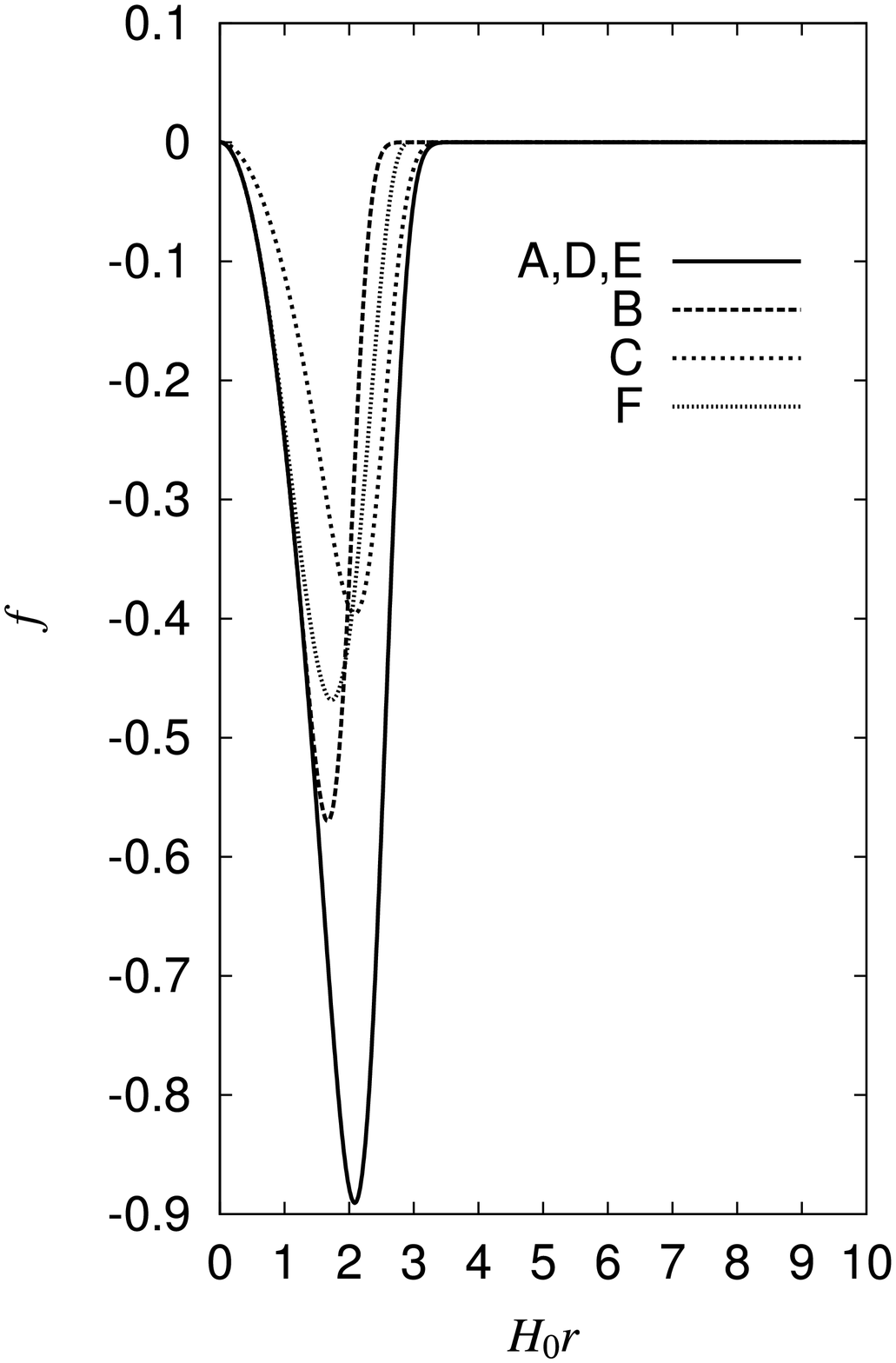}~
(b)\includegraphics[scale=0.5]{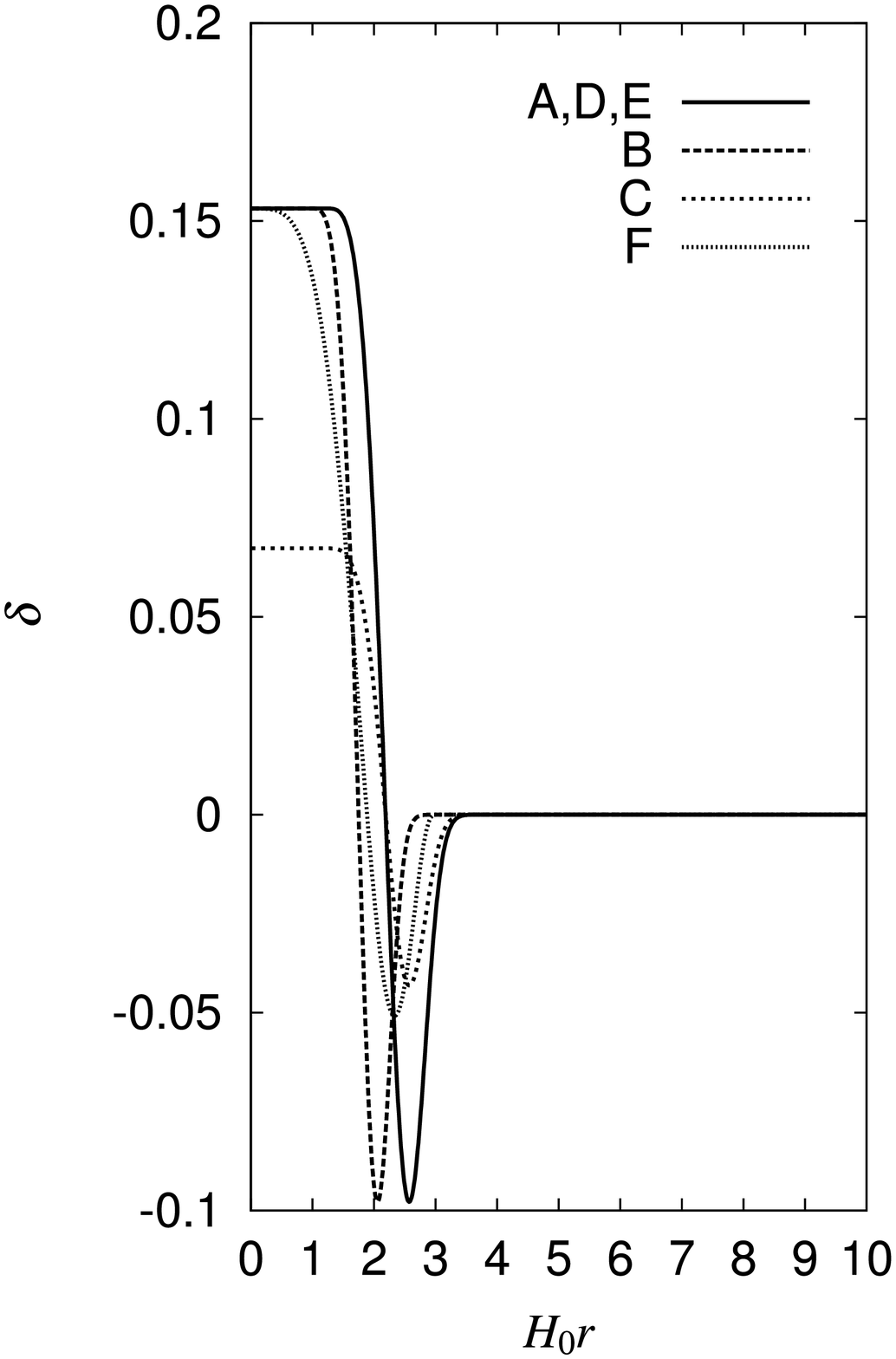}\\
FIG.1
\end{figure}
\begin{figure}[htbp]
(a)\includegraphics[scale=0.4]{2a}~
(b)\includegraphics[scale=0.4]{2b}
(c)\includegraphics[scale=0.4]{2c}~
(d)\includegraphics[scale=0.4]{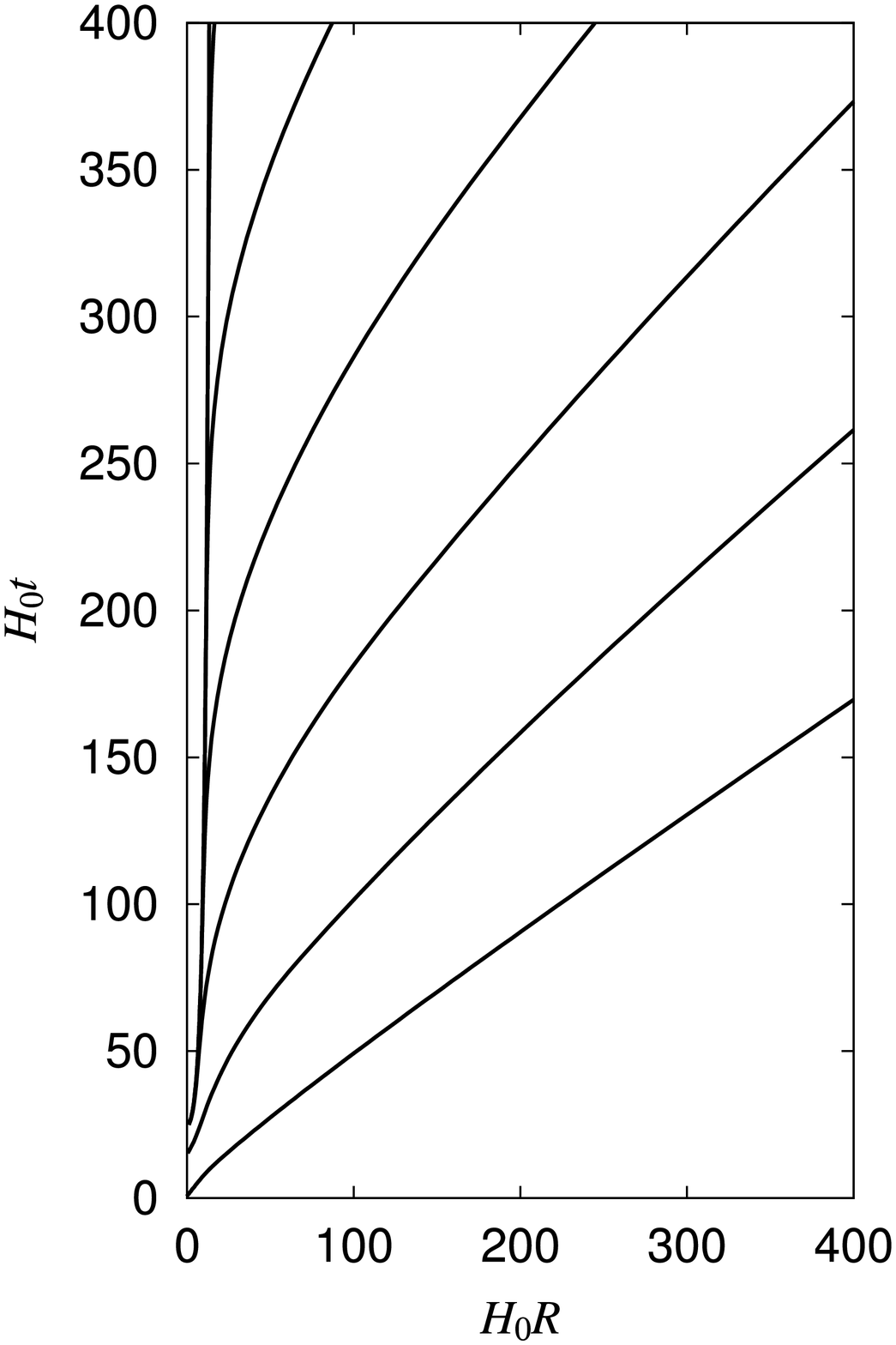}\\
FIG.2
\end{figure}
\begin{figure}[htbp]
\includegraphics[scale=0.8]{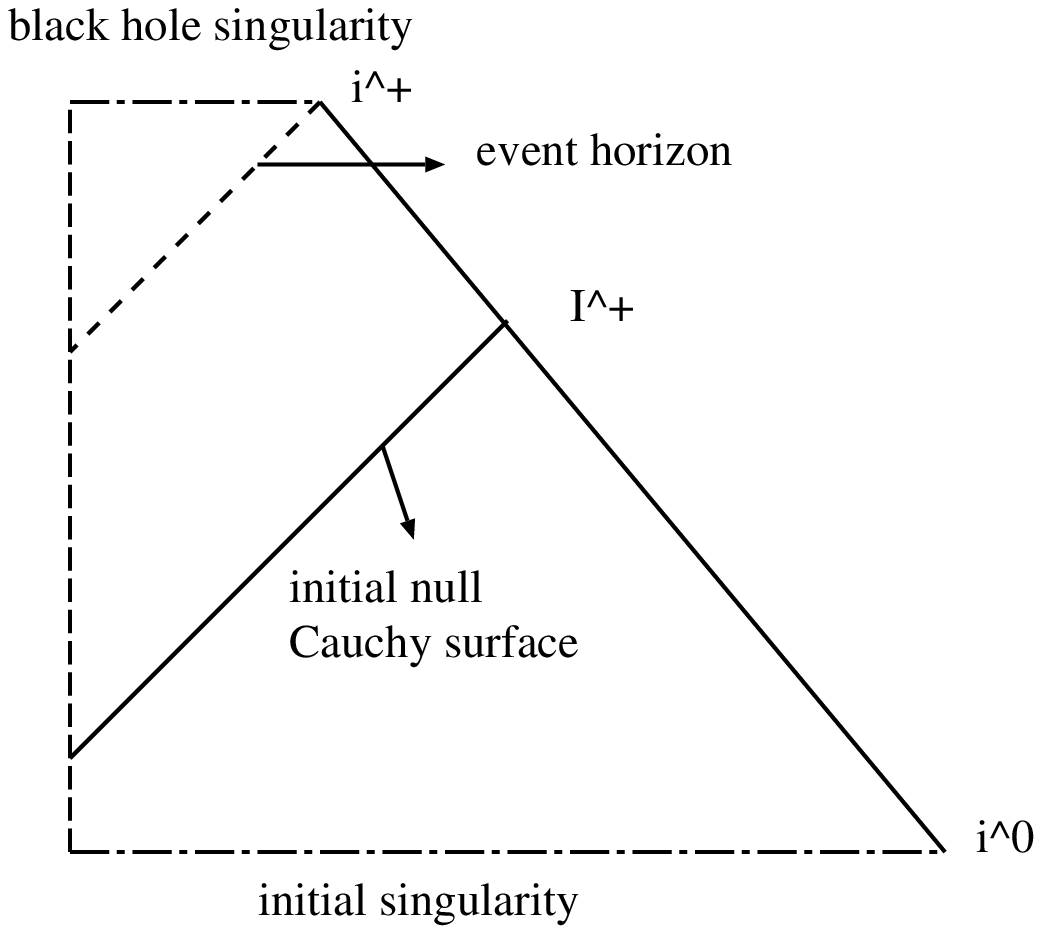}\\
FIG.3
\end{figure}
\begin{figure}[htbp]
\includegraphics[scale=0.4]{4}\\
FIG.4
\end{figure}
\begin{figure}[htbp]
(a)\includegraphics[scale=0.4]{5a}~
(b)\includegraphics[scale=0.4]{5b}
(c)\includegraphics[scale=0.4]{5c}~
(d)\includegraphics[scale=0.4]{5d}\\
FIG.5
\end{figure}
\begin{figure}
(e)\includegraphics[scale=0.4]{5e}~
(f)\includegraphics[scale=0.4]{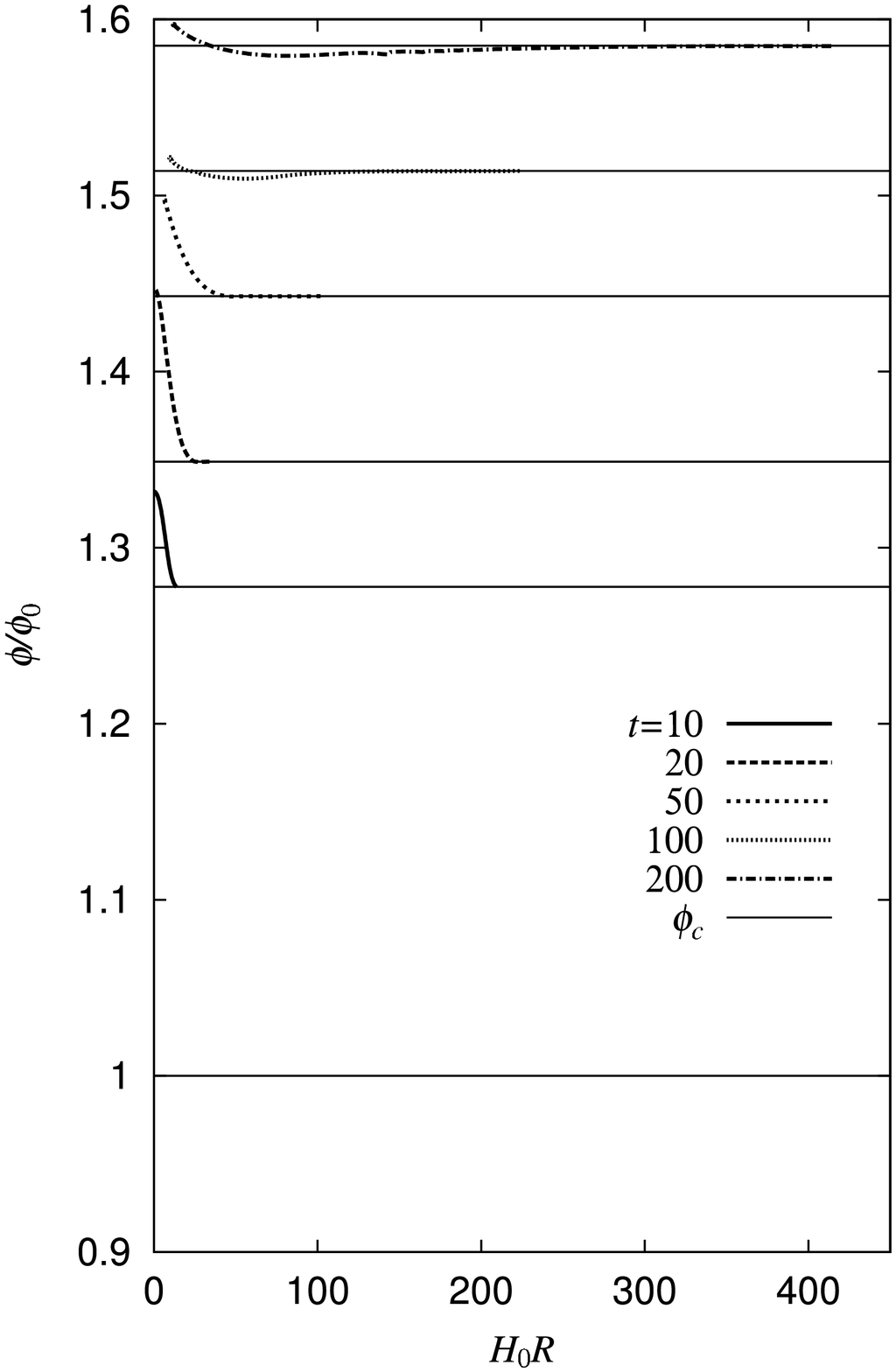}\\
FIG.5 (continued)
\end{figure}
\begin{figure}[htbp]
(a)\includegraphics[scale=0.4]{6a}~
(b)\includegraphics[scale=0.4]{6b}
(c)\includegraphics[scale=0.4]{6c}~
(d)\includegraphics[scale=0.4]{6d}\\
FIG.6
\end{figure}
\begin{figure}
(e)\includegraphics[scale=0.4]{6e}~
(f)\includegraphics[scale=0.4]{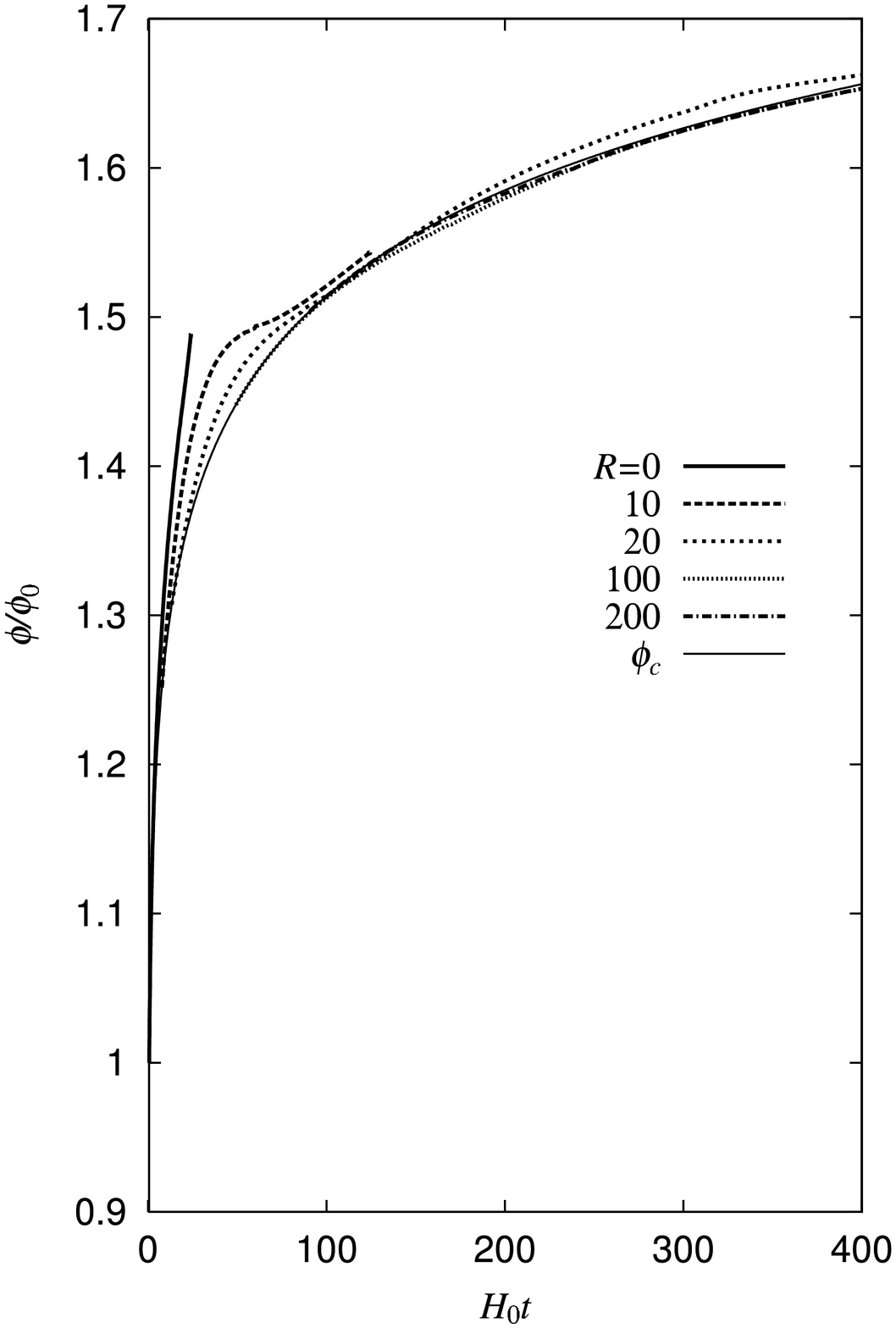}\\
FIG.6 (continued)
\end{figure}
\begin{figure}[htbp]
(a)\includegraphics[scale=0.4]{7a}~
(b)\includegraphics[scale=0.4]{7b}
(c)\includegraphics[scale=0.4]{7c}~
(d)\includegraphics[scale=0.4]{7d}\\
FIG.7
\end{figure}
\begin{figure}
(e)\includegraphics[scale=0.4]{7e}~
(f)\includegraphics[scale=0.4]{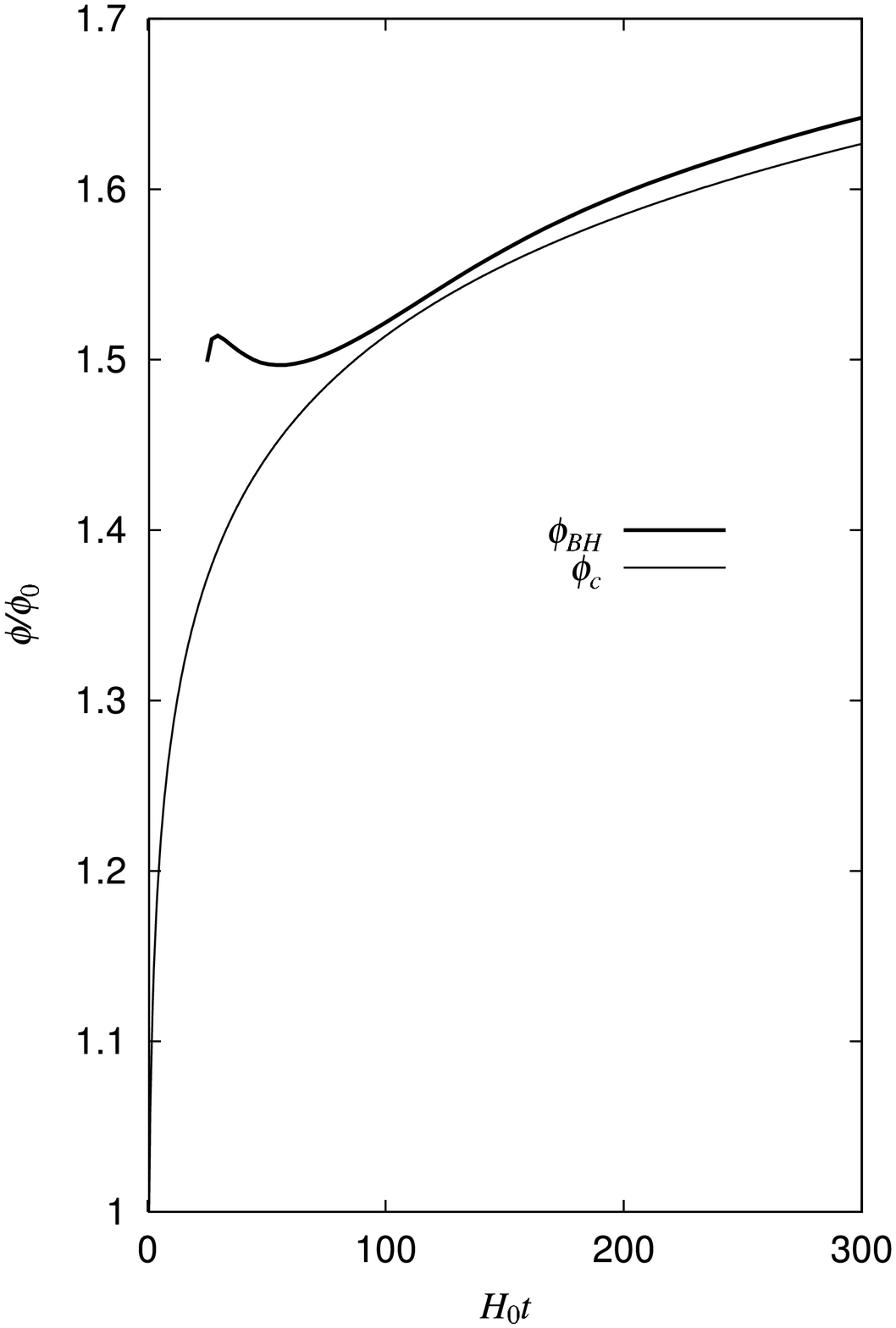}\\
FIG.7 (continued)
\end{figure}

\end{document}